\documentclass[aps,prl,twocolumn,superscriptaddress,groupedaddress]{revtex4-1}  
\usepackage{graphicx}  
\usepackage{bm}        
\usepackage{amssymb}   
\usepackage{amsmath}
\usepackage{hyperref}
\usepackage{natbib}
\hypersetup{
	colorlinks=true,
	urlcolor= blue,
	citecolor=blue,
	linkcolor= blue,
	bookmarks=true,
	bookmarksopen=false,
}

\begin{document}

\title{Topological Theory for Perfect Metasurface Isolators}
\author{Wai Chun Wong} \affiliation{Department of Applied Physics, The Hong Kong Polytechnic University, Hong Kong, China}
\author{Wenyan Wang} \affiliation{Department of Applied Physics, The Hong Kong Polytechnic University, Hong Kong, China}
\author{Wang Tat Yau} \affiliation{Department of Applied Physics, The Hong Kong Polytechnic University, Hong Kong, China}
\author{Kin Hung Fung} \email{khfung@polyu.edu.hk}\affiliation{Department of Applied Physics, The Hong Kong Polytechnic University, Hong Kong, China}
\date{\today}

\begin{abstract}
	
We introduce topological theory of perfect isolation: perfect transmission from one side and total reflection from another side simultaneously. The theory provides an efficient approach for determining whether such a perfect isolation point exists within a finite parameter space. Herein, we demonstrate the theory using an example of a Lorentz non-reciprocal metasurface composed of dimer unit cells. Our theory also suggests that perfect isolation points can annihilate each other through the coalescence of opposite topological charges. Our findings could lead to novel designs for high-performance optical isolators.
	
\end{abstract}

\maketitle
Recently, various types of topological theories have been applied successfully to the design of backscattering-immune one-way photonic systems \cite{wang2009observation,wang2008reflection,khanikaev2013photonic,PhysRevLett.106.093903}. For example, topological band theory can be used to design robust one-way waveguides that guide waves through the topological edge modes of photonic crystals \cite{lu2014topological}. Topological approaches have also been applied to study numerous novel and exotic effects in reciprocal systems, such as bound states in continuum \cite{PhysRevLett.113.257401,PhysRevLett.118.267401,PhysRevA.90.053801,doeleman2018experimental}, complete polarization conversion \cite{PhysRevLett.119.167401}, and coherent perfect reflection \cite{PhysRevB.98.081405}. These approaches entail considering vertices or singularities in some vector fields and identifying the vertex or singularity with notable phenomena. For example, the singularity of the polarization direction of far-field radiation could correspond to bound states in continuum, vertices of complex reflection or transmission coefficients could corresponds to complete polarization conversion or coherent perfect reflection. Such vertices or singularities are called topological charges because they could be associated with a topological invariant, such as winding number.


Lorentz non-reciprocal devices are crucial to photonics applications because such devices could stabilize laser operation by suppressing backward reflections and enlarge the design space of an optical communication system. Various designs of Lorentz non-reciprocal devices have been proposed by previous studies \cite{PhysRevLett.109.033901,yu2009complete,PhysRevB.84.075477,fang2012realizing,Wang:05}. Nevertheless, designs based on photonic crystals and other approaches are often bulky \cite{PhysRevB.84.075477,fang2012realizing,Wang:05}. For example, a Faraday rotator \cite{Aplet:64}, which utilizes non-reciprocal polarization rotation in magneto-optic (MO) materials, could achieve non-reciprocal transmission only when a fairly large propagation distance in the MO materials is provided. The number of designs for miniaturization, such as metasurface isolators, is limited because of the high complexity and difficulty in determining a systematic scheme for optimization.

In this paper, we introduce topological theory for perfect isolating effect: perfect transmission from one side and total reflection from the other side simultaneously. The perfect isolation phenomena in our examples correspond to the zero of a real vector in the optimization parameter space. For a theoretical demonstration, we introduce a Lorentz non-reciprocal metasurface compose of dimer unit cells. Furthermore, we can observe the annihilation of these perfect isolation points though changing parameters. This work is expected to open up new avenues for robust ultra-high-isolation non-reciprocal metasurfaces and provide a novel approach to the design of perfect metasurface isolators.

\begin{figure*}[!tbp]
	\includegraphics[width=5.25in]{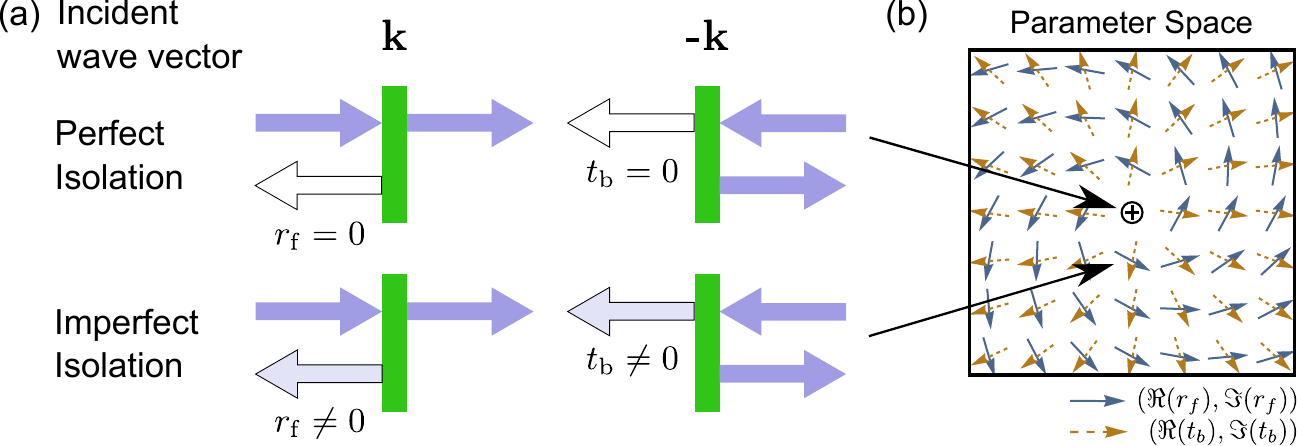}
	\centering
	\caption {\small (a) Schematic of the difference between perfect isolation and imperfect isolation. Each green slab represents a Lorentz non-reciprocal surface. Blue arrows represent the incident, reflected, or transmitted plane waves. One must have $r_\mathrm{f}=0$ and $t_\mathrm{b}=0$ for a perfect isolation case and one must have $r_\mathrm{f}\neq0$ or $t_\mathrm{b}\neq0$ for an imperfect isolation case. (b) Schematic of vector $\bm{F}$ by Eq.\ (\ref{winding}) in parameter space. Each blue vector represents the vector $\left(\Re(r_{f}),\Im(r_{f})\right)$ and each dashed yellow vector represents the vector $\left(\Re(t_{b}),\Im(t_{b})\right)$. Perfect isolation corresponds to zero of $\bm{F}$. Imperfect isolation corresponds to a regular vector. }
	\label{schme}
\end{figure*}


First, we consider a Lorentz non-reciprocal surface and study the transmission and reflection properties when a plane wave with wave vectors \textbf{k} and \textbf{-k} is incident on this surface. Let $t_\mathrm{f}$ and $r_\mathrm{f}$ represent the transmission coefficient and reflection coefficient for waves with wavevector \textbf{k}. Similarly, the coefficients for waves with wavevector \textbf{-k} are denoted as $t_\mathrm{b}$ and $r_\mathrm{b}$. We define the perfect isolation condition as $|t_\mathrm{f}|^2=1$ and $|t_\mathrm{b}|^2=0$, satisfying perfect transmission from one side and no transmission from the other side simultaneously [Fig.\ \ref{schme}(a)]. If the surface has no absorption or other channels that could remove energy (e.g. diffraction channel), conservation of energy ensure that $|t_\mathrm{f}|^2+|r_\mathrm{f}|^2=1$; therefore we could translate the conditions to $r_\mathrm{f}=0$ and $t_\mathrm{b}=0$. 
To discuss the topology of these two zeros, we represent $r_\mathrm{f}$ and $t_\mathrm{b}$ as a four-dimensional vector $\bm{F}$:
\begin{equation}
\bm{F}=\left(F_{1}, F_{2}, F_{3}, F_{4}\right)=\left(\Re(r_{f}),\Im(r_{f}),\Re(t_{b}),\Im(t_{b})\right).
\end{equation}
Where $\Re(z)$ and $\Im(z)$ are the real part and the imaginary part of $z$, respectively; that is, $\bm{F}$ is composed of the real and imaginary parts of $r_\mathrm{r}$ and $t_\mathrm{b}$. As illustrated in Fig.\ \ref{schme}(b), a perfect isolation condition corresponds to $\bm{F}=0$ and an imperfect isolation condition corresponds to a regular vector $\bm{F}\neq0$. Mathematically, a topological invariant (winding number, denoted as $W$) could be assign to zero of $\bm{F}$ as follow\cite{dubrovin2012modern}:
\begin{equation}
W=\frac{1}{4\pi^2}\oint_{\Omega} \dfrac{1}{|\bm{F}|^2}\det\begin{pmatrix}
F_{1} & F_{2} & F_{3} & F_{4} \\[3pt]
\frac{\partial F_{1}}{\partial s} & \frac{\partial F_{2}}{\partial s} & \frac{\partial F_{3}}{\partial s} & \frac{\partial F_{4}}{\partial s} \\[3pt] 
\frac{\partial F_{1}}{\partial t} & \frac{\partial F_{2}}{\partial t} & \frac{\partial F_{3}}{\partial t} & \frac{\partial F_{4}}{\partial t} \\[3pt] 
\frac{\partial F_{1}}{\partial u} & \frac{\partial F_{2}}{\partial u} & \frac{\partial F_{3}}{\partial u} & \frac{\partial F_{4}}{\partial u} 
\end{pmatrix} \,ds\,dt\,du. 
\label{winding}
\end{equation}
Where  $|\bm{F}|=\sqrt{F_{1}^2+F_{2}^2+F_{3}^2+F_{4}^2}$; $\Omega$ is a three-dimensional closed surface surrounding the perfect isolation point in the parameter space, and $(s,t,u)$ is a set of the parametrization variables of $\Omega$. The Integration in Eq.\ (\ref{winding}) can be used to calculate the higher dimensional solid angle which the vector $\bm{F}$ sweeps around the isolation point.
Notably, this definition is highly related to the notion of winding number used in periodically driven two-dimensional (2D) systems \cite{PhysRevX.3.031005,Maczewsky2017}, and to the generalized winding number of 2D vector fields used in previous studies \cite{PhysRevLett.113.257401, PhysRevLett.119.167401,PhysRevB.98.081405}. Because perfect isolation is topological, our theory predict this effect have other topological protected properties such as robust against small perturbations and annihilations between opposite charges.

\begin{figure}[!tbp]
	\includegraphics[width=3.2in]{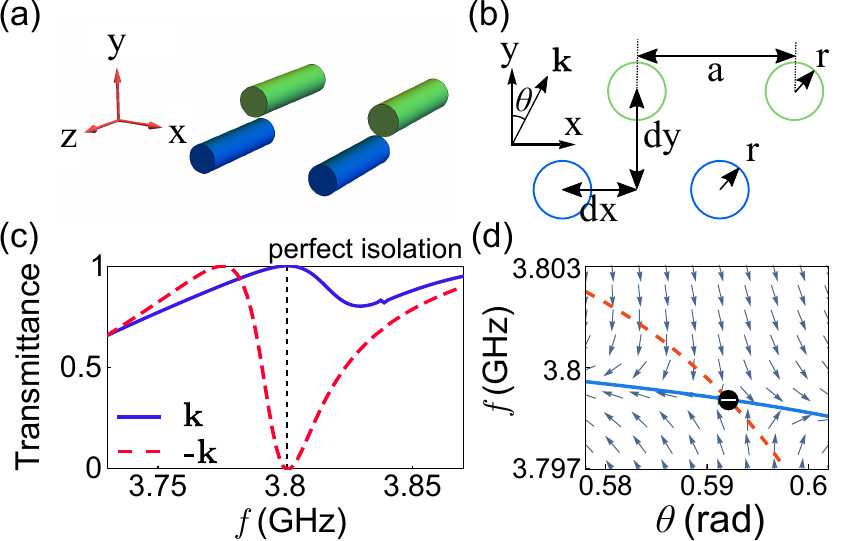}
	\centering
	\caption {\small (color online) (a), (b) Schematic of Lorentz non-reciprocal grating with dimer unit cells. Two cylinder layers are arranged such that vertical separation is $dy$ and horizontal offset is $dx$; that is, center-to-center displacement between the different layers is $\textbf{dr}=\left(dx,dy\right)$.  Ferromagnetic YIG cylinders are indicated in green and dielectric cylinders are indicated in blue. (c) Transmittance spectra of grating in (a) and (b) with $a=50$ mm, $dx=12$ mm, $dy=14.72$ mm, $r=1$ mm, $n=10.20$, and $\theta=0.5735$. The blue Solid line and red dashed line correspond to the incident wave vectors $\textbf{k}$ (i.e., forward propagation) and $-\textbf{k}$ (i.e. backward propagation). (d) Vector field $(\text{Re}(\tilde{r}_{\mathrm{f}}),\text{Re}(\tilde{t}_{\mathrm{b}}))$ with the same parameters as in (c), but Eq.\ (\ref{a},\ref{b}) are assumed such that $\text{Im}(\tilde{r}_{\mathrm{f}})=\text{Im}(\tilde{t}_{\mathrm{b}})=0$.}
	\label{fig1}
\end{figure}

To demonstrate our theory, we consider a Lorentz nonreciprocal grating with two cylinder layers [Fig.\ \ref{fig1}(a) and \ref{fig1}(b)]. Each unit cell consists of two cylinder composed of different materials. The first cylinder is a lossless dielectric cylinder with refractive index $n$ and with $ \text{Im}(n) = 0$. A later section of the present paper explains the effect of loss. The second cylinder is composed of a ferromagnetic yttrium-iron garnet (YIG). When an external static magnetic field is applied in the z-direction, the ferromagnetic YIG breaks Lorentz reciprocity and is characterized by permittivity $\epsilon = 15 \epsilon_{0}$ and magnetic permeability
\begin{equation}
\bar{\mu} =
\begin{pmatrix}
\mu 		& i \Delta 	& 0 		\\
-i \Delta 	& \mu 		& 0			\\
0 			& 0 		& 1
\end{pmatrix}\,.
\end{equation}
Where $\mu = 1+\omega_{m}\omega_{h}/(\omega_{h}^{2}-\omega^{2})$ and $\Delta = -\omega_{m}\omega/(\omega_{h}^{2}-\omega^{2})$, with precession frequency $\omega_{h}=\gamma H_{0} $, gyromagnetic ratio $\gamma$, applied static magnetic field $H_{0}$, and characteristic frequency $\omega_{m} = \gamma M_{s}$. $M_{s}$ represents the saturation magnetization in the ferromagnetic materials. Herein, we set $M_{s}$ to $1750$ Oe and $H_{0}$ to $500$ Oe.

To evaluate transmittance and reflectance of this grating, we apply finite-element Method (FEM) using \textsc{COMSOL Multiphysics}. We confine $\textbf{k}$ to the $x$-$y$ plane and consider only TE polarization (${\hat{\textbf{E}}}=\hat{z}$) because TM polarization waves do not have non-reciprocal properties. Fig.\ \ref{fig1}(c) illustrates the $T_{\mathrm{f}}$ and $T_{\mathrm{b}}$ spectra of the grating with specific incident angle $\theta$. All parameters used in this calculation are described in the caption of Fig.\ \ref{fig1}. We observe that this surface can support non-reciprocal transmission over a wide frequency range. The transmittance difference $\Delta T=T_{\mathrm{f}}-T_{\mathrm{b}}$ can reach $99.9996\%$ at a frequency $f$ of $3.801$ GHz, which is near-perfect isolation. According to these parameters, no diffraction will occur because $|k_x\pm2\pi/a| >k_0$; therefore, we could apply our theory to this isolation point. 

\begin{figure*}[!htbp]
	\includegraphics[width=6.5in]{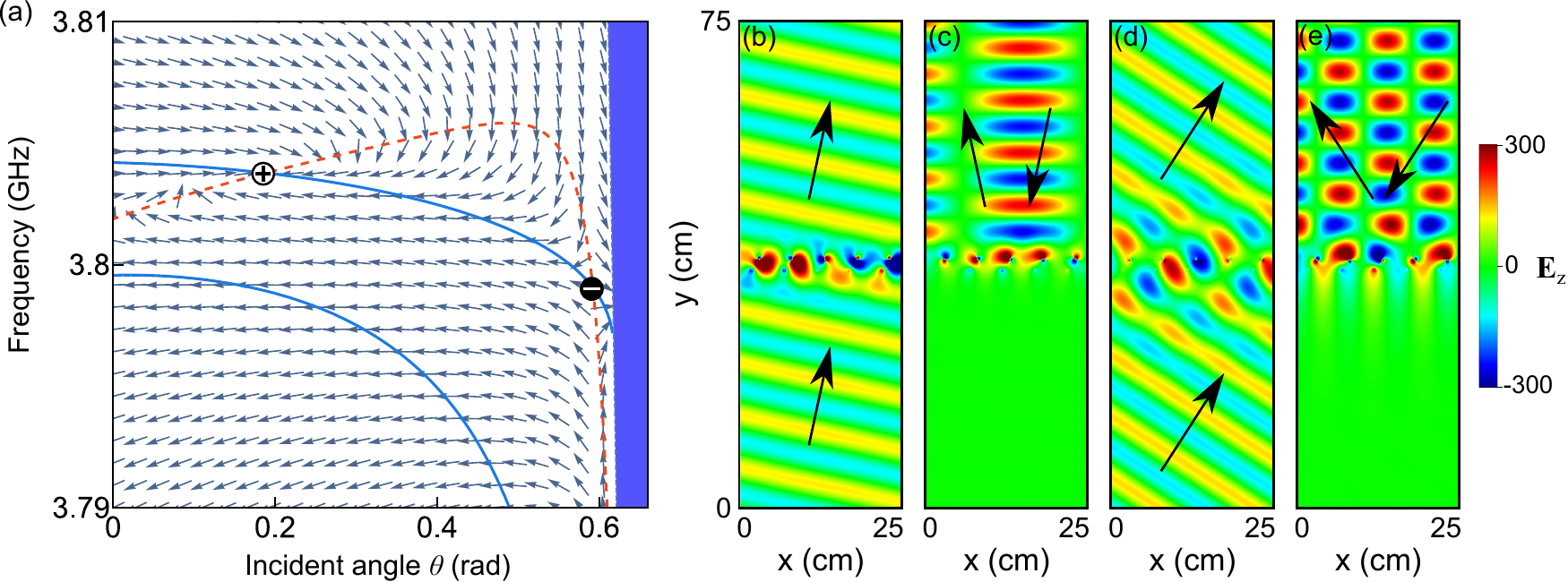}
	\centering
	\caption{\small (color online) (a) Vector field $(\text{Re}(\tilde{r}_{\mathrm{f}}),\text{Re}(\tilde{t}_{\mathrm{b}}))$ defined in Eq. (4-7) when the configuration is the same as in Fig.\ \ref{fig1}(c); $b^{0}_{\mathrm{die}}$ and $d_{y}$ are assumed to be $b^{1}_{\mathrm{YIG}}$ and $(\pi/2-\theta)/k_{y}$ such that $\text{Im}(\tilde{r}_{\mathrm{f}})=\text{Im}(\tilde{t}_{\mathrm{b}})=0$, and the vectors are normed individually. Two nodal lines $\text{Re}(\tilde{r}_{\mathrm{f}})=0$ and $\text{Re}(\tilde{t}_{\mathrm{b}})=0$ are graphed as a dashed red line and as a solid blue line, respectively; therefore, the topological point could also be indicated by crossing of the two nodal lines. Shaded region indicates that diffraction will occur. (b)-(e) FEM-simulated electric field of the grating at topological charges in (a) under incident plane waves. Arrows denote the directions of incident waves and transmitted or reflected waves. For (b) and (c), the parameters used are $r=1$ mm, $a=50$ mm, $dx=12$ mm, $dy=17.10$ mm, $n=11.42$, $\theta=0.2133$, and $f=3.804$ GHz, which correspond to positive charge with lower $\theta$. For (d) and (e), the parameters used are $r=1$ mm, $a=50$ mm, $dx=12$ mm, $dy=14.72$ mm, $r=1$ mm, $n=10.20$, $\theta=0.5735$, and $f=3.801$ GHz, which correspond to negative charge with higher $\theta$.}
	\label{fig2}
\end{figure*}

Here, we calculate the winding number for this example.  We set $\Omega$ as a 3-sphere $S^3$ embedded in $\left(\theta, dy, n, f\right)$ with center at the parameter explained in Fig.\ \ref{fig1}(c), and set the other parameters constant. Through numerical calculations, we can determine that the perfect isolation point in the previous example is topological with winding number $-1$. This nonzero winding number ensures that a perfect isolation point exist inside the 3-sphere $S^3$.

To further study the conditions for perfect isolation, we formulate a 2$\times$2 matrix model based on multiple scattering theory. In general, in multiple scatting theory, electromagnetic fields near cylinders are expanded into sums of cylindrical harmonics and the couplings between all cylinders are considered. The formulation process is detailed in the Supplemental Material. Although all harmonic orders contribute to any scattering process, here, we model each cylinder by its dominant harmonic, which is $+1$ order for ferromagnetic YIG cylinder and $0$ for dielectric cylinder. This choice is motivated by comparing magnitude of different Mie's coefficient and also motivated by observing the electric field pattern in Fig.\ \ref{fig2}(b-e). We present a comparison of the forward and backward transmittances obtained by FEM and by the aforementioned model in the Supplemental Material. We reveal that these two methods have high agreement with little discrepancy; therefore, we can regard this approximation as a small perturbation and that the topology is not altered. 
On the basis of this simplification, forward reflection and backward transmission coefficients can be calculated as follow \cite{yasumoto2005electromagnetic}:
\begin{align}
r_{\mathrm{f}}=& \frac{2e^{i\theta}}{a k_{y}((b_{\mathrm{die}}^{-1}-L)(b_{\mathrm{YIG}}^{-1}-L)-\zeta_{+}\xi_{+})} \tilde{r}_{\mathrm{f}}\label{tilde}\\
t_{\mathrm{b}}=& \frac{1}{((b_{\mathrm{die}}^{-1}-L)(b_{\mathrm{YIG}}^{-1}-L)-\zeta_{-}\xi_{-})} \tilde{t}_{\mathrm{b}}
\label{tilde2}
\end{align}
where
\begin{align}
\begin{split}
\tilde{r}_{\mathrm{f}} = & (b_{\mathrm{die}}^{-1}-L)e^{i(\theta-k_{y}d_{y})}-(b_{\mathrm{YIG}}^{-1}-L)e^{-i(\theta-k_{y}d_{y})}\\
&+\xi_{+} e^{-ik_{x}d_{x}}-\zeta_{+} e^{ik_{x}d_{x}}\label{trf}
\end{split}\\
\begin{split}
\tilde{t}_{\mathrm{b}} = & (L-b_{\mathrm{die}}^{-1}+\frac{2}{ak_{y}})(L-b_{\mathrm{YIG}}^{-1}+\frac{2}{ak_{y}})\\
&-\left(\zeta_{-}+\frac{2}{ak_{y}}e^{-i\phi}\right)\left(\xi_{-}+\frac{2}{ak_{y}}e^{i\phi}\right)\label{ttb}
\end{split}.
\end{align}

Here $\phi=\theta+k_{y}d_{y}-k_{x}d_{x}$; $b_{\mathrm{die}}$ and $b_{\mathrm{YIG}}$ are the dominant Mie's coefficient of the dielectric and ferromagnetic YIG cylinders, respectively; $L=\sum_{l=1}^{\infty}H^{(1)}_{0}(lk_0a)(e^{ilk_xa}+(-1)^{n-m}e^{-ilk_xa})$ is the lattice sum of the grating; and $\zeta_{+}=\sum_{l=-\infty}^{\infty}H^{(1)}_{1}(k_0|la\bm{\hat{x}}+\textbf{dr}|)e^{i(lh k_x+\arg(la\bm{\hat{x}}+\textbf{dr}))}$ and  $\xi_{+}=\sum_{l=-\infty}^{\infty}H^{(1)}_{-1}(k_0|la\bm{\hat{x}}-\textbf{dr}|)e^{i(lh k_x-\arg(la\bm{\hat{x}}-\textbf{dr}))}$ are the relative lattice sums between the dielectric and ferromagnetic YIG cylinders, respectively. $\arg(\textbf{r})$ is the polar angle of $\textbf{r}$. $\zeta_{-}$ and $\xi_{-}$ is  $\zeta_{+}$ and $\xi_{+}$ when $\textbf{k}$ is reversed.

\begin{figure*}[!htbp]
	\includegraphics[width=5.25in]{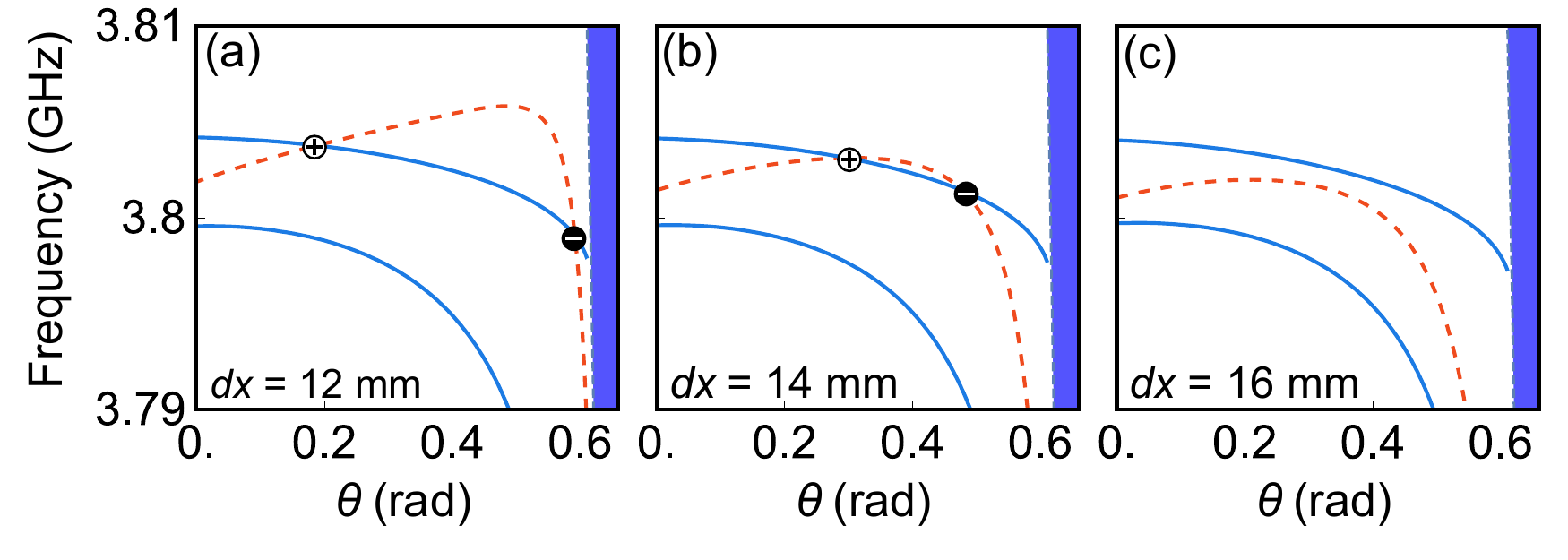}
	\centering
	\caption{\small (color online) Evolution of topological charges in the parameter space. (a) Nodal lines and topological charge when $dx=12$ mm. Other parameters are the same as in Fig.\ \ref{fig1}(c). $b^{0}_{\mathrm{die}}$ and $d_{y}$ are assumed to be $b^{1}_{\mathrm{YIG}}$ and $(\pi/2-\theta)/k_{y}$ such that $\text{Im}(\tilde{r}_{\mathrm{f}})=\text{Im}(\tilde{t}_{\mathrm{b}})=0$ according to Eq. (\ref{a},\ref{b}). Two nodal lines $\text{Re}(\tilde{r}_{\mathrm{f}})=0$ and $\text{Re}(\tilde{t}_{\mathrm{b}})=0$ are shown as a dashed red line and as a solid blue line, respectively. Both (b) and (c) are essentially similar to (a) but with $dx$ increased to 14 and 16 mm.}
	\label{fig4}
\end{figure*}

 Because $\tilde{r}_{\mathrm{f}}$, $\tilde{t}_{\mathrm{b}}$ and $r_{\mathrm{f}}$, $t_{\mathrm{b}}$ are related by Eq.\ (\ref{tilde},\ref{tilde2}), that is multiplied a non-zero continuous factor, $\bm{\tilde{F}}=\left(\Re(\tilde{r}_{\mathrm{f}}),\Im(\tilde{r}_{\mathrm{f}}),\Re(\tilde{t}_{\mathrm{b}}),\Im(\tilde{t}_{\mathrm{b}})\right)$ is a continuously transformed version of $\bm{F}$. Moreover, both $\bm{\tilde{F}}$ and $\bm{F}$ have the same topology. This transformation may appear unnecessary, but $\bm{\tilde{F}}$ has superior analytical properties to $\bm{F}$ as detailed in the following explanation: First, we have two useful identities about the lattice and relative lattice sums \cite{Botten1,Botten2}: $\Re{(L)}=1-\frac{2}{hk_y}$, $\zeta_\pm+ \bar{\xi}_\pm=\mp \frac{4i}{hk_y} \sin(\theta\mp k_y d_y) e^{\mp ik_xd_x}$, where $k_y, k_x>0$ for both directions and $\bar{\xi}$ is the complex conjugate of $\xi$. Substituting the two identities into Eq.\ (\ref{trf}-\ref{ttb}) reveals sufficient conditions for $\text{Im}(\tilde{r}_{\mathrm{f}})$ and $\text{Im}(\tilde{t}_{\mathrm{b}})$ to become equal to zero:
\begin{align}
b_{\mathrm{die}} = b_{\mathrm{YIG}} &\implies \text{Im}(\tilde{r}_{\mathrm{f}})=0 \label{a} \\
k_{y} d_{y}+\theta = \frac{\pi}{2} &\implies \text{Im}(\tilde{t}_{\mathrm{b}})=0. \label{b}
\end{align}
The Mie's coefficients $b$ represents the amplitudes of the fields scattered by the cylinder array. The first condition can be understood as equating the amplitudes of the scattered field by two cylinder arrays, which could lead to destructive interference and could result in a situation with no reflected waves, that is, a reflected wave with an amplitude of zero. 


Using the conditions set in Eq.\ (\ref{a}-\ref{b}), we further investigate the topology of $\bm{\tilde{F}}$. Fig.\ \ref{fig1}(d),\ref{fig2}(a) shows the vector field $(\text{Re}(\tilde{r}_{\mathrm{f}}),\text{Re}(\tilde{t}_{\mathrm{b}}))$. The two conditions in Eq.\ (\ref{a}-\ref{b}) are assumed such that the nonzero components of $\bm{\tilde{F}}$ are $\text{Re}(\tilde{r}_{\mathrm{f}}),\text{Re}(\tilde{t}_{\mathrm{b}})$ only and are shown in the 2D vector field. In both figures, all the vectors are normalized individually.
The vector field in Fig.\ \ref{fig1}(d),\ref{fig2}(a) can be considered the projection of $\bm{\tilde{F}}$ in the parameter space along a surface that is defined by Eq.\ (\ref{a},\ref{b}). 
This specific surface is chosen because vectors on this surface are oriented in the 2D plane such that no essential information is lost when the vectors are projected.
Figure \ref{fig1}(d) shows that the perfect isolation point described in Fig.\ \ref{fig1} can be associated with a topological charge using projection.

Figure \ref{fig2}(a) illustrates two topological charges. The negative charge with higher $\theta$ corresponds to the example in Fig.\ \ref{fig1}, whereas the other positive charge is explained later in this paper. The negative charge is slightly shifted when compare with the results calculated by FEM. This shift is due to the omission of other cylindrical harmonics. Contributions from other non-dominant harmonics could be treated as small perturbations. Because of the robustness of topological charges, the perfect isolation point is preserved. We have also verify that if more harmonics are included, the location of the perfect isolation pointcan exhibit higher agreement with the FEM result. For the positive charge with lower $\theta$, we confirm that there exists near-perfect isolation near the topological charge, that is, we can estimate that $t_{\mathrm{b}}, r_{\mathrm{f}} =0$ through the FEM and multiple scattering method. Fig.\ \ref{fig2}(b-e) shows the corresponding electric field patterns for the forward and backward incident waves at both topological charges.

Here, we investigate the evolution of isolation points with changes in $dx$. With a small increase in $dx$, the two topological charges move toward each other [Fig.\ \ref{fig4}(b)]. As $dx$ increases further, the two opposite charges collide and annihilate each other. In other words, the nodal line $\text{Re}(\tilde{r}_{\mathrm{f}})=0$ (dashed red) move down as $dx$ increases, but the other nodal line $\text{Re}(\tilde{t}_{b})=0$ (solid blue) is less sensitive to the small change in $dx$; eventually, the two interception points meet. After the annihilation, no interception point (i.e., no topological charge) exists [Fig.\ \ref{fig4}(c)]. This is a generic situation of annihilation between two topological charges \cite{PhysRevLett.113.257401,PhysRevLett.118.267401}.

In the preceding paragraphs, we consider cylinders with no absorption loss; here, we discuss the effect of loss on isolation points. In a lossy system, no principle can guarantee that the sum of transmittance $T$ and reflectance $R$ will be unity. Therefore, we cannot relate perfect isolation to the topological point. Nevertheless, the point where $r_{\mathrm{f}}=0$ and $t_{\mathrm{b}}=0$ is still topological even with small absorption loss because the point is indicated to be a singularity  in Eq.\ (\ref{winding}). At this point, the scattering properties of the grating are such that no reflection is available from one side and no transmission is possible from the other side. These properties can still prevent backward transmission of energy. In the Supplemental Material, we also present the results using lossy dielectric. We successfully trace the topological point in Fig.\ \ref{fig1} with an increase in absorption loss and provide a lossy system with $r_{\mathrm{f}}, t_{\mathrm{b}}=0$ with high transmittance difference.

In summary, we introduce topological theory describing perfect isolation phenomena as singularities in the parameter space. We determent that a winding number can be associated with each singularity through Eq.\ (\ref{winding}). Our theory predicts the robustness and annihilation between two charges as a natural consequence of perfect isolation being topological. To demonstrate our theory, we introduce a Lorentz non-reciprocal dimer metasurface with dielectric and ferromagnetic materials. With an external static magnetic field, this metasurface exhibits strong non-reciprocal transmissions. Under specific parameters, the grating supports perfect isolation, and this perfect isolation is topological with winding number $-1$. Furthermore, the topological points are investigated using an analytical model based on multiple scattering theory. We demonstrate that, through the tuning of geometric parameters, two isolation points with opposite charge can annihilate each other. 

We thank Yongliang Zhang, Jin Wang, Kai Fung Lee, Jensen Li, and C.T. Chan for their assistance and fruitful discussions. This work was supported by the Hong Kong Research Grants Council through project nos. C6013-18G and 15301917.

\bibliography{ref}
\end{document}